\documentstyle[11pt,fleqn,epsf]{article}
\def\ref{par\noindent\hangindent=6mm\hangafter=1}
\begin{document}
\vbox{
\rightline{Nuovo Cimento B 113 (May 1998) 677-682}
\rightline{gr-qc/9610018}
}
\baselineskip 8mm
%
\begin{center}
{\bf Double Darboux method for the Taub continuum}

\bigskip

H. Rosu\footnote{E-mail: rosu@ifug3.ugto.mx} 
and J. Socorro\footnote{E-mail: socorro@ifug4.ugto.mx} 

{\it Instituto de F\'{\i}sica de la Universidad de Guanajuato, Apdo Postal
E-143, Le\'on, Gto, M\'exico}


\end{center}

\bigskip
\bigskip


{\bf Summary}. - The strictly
isospectral double Darboux method is applied to the quantum
Taub model in order to generate a one-parameter family of strictly
isospectral potentials for this case. The family we build is based on a
scattering Wheeler-DeWitt solution first discussed by
Ryan and collaborators that we slightly modified according to a suggestion
due to Dunster.
The strictly isospectral Taub potentials possess
different (attenuated) scattering states with respect to the original
Taub potential.



\vskip 1cm



Quantum cosmology and its supersymmetric extension \cite{Mo} are
an interesting ``laboratory" for techniques of
current use in nonrelativistic quantum mechanics. One such technique is the
strictly isospectral double Darboux method (SIDDM), which is a procedure within
Witten's supersymmetric quantum mechanics \cite{cks}.

Previously, we have applied SIDDM to closed, radiation-filled
Friedmann-Robertson-Walker (FRW) quantum universes, obtaining a one-parameter
family of strictly isospectral FRW quantum potentials and the corresponding
wavefunctions \cite{rs96}. The Taub minisuperspace model is a separable
quantum problem \cite{ry} and therefore is well suited for SIDDM.
It is our purpose in this paper to develop the method for
the continuous part of the spectrum of the Taub model. The motivation for
doing this exercise was found in a paper by Sukhatme and collaborators
\cite{suk}, who obtained bound states in the continuum in quantum mechanics by
SIDDM. However, for the cosmological Taub case the method does not
produce such states as we shall show in the following.
 

SIDDM is a technique of deleting followed
by reinstating an energy level of a one-dimensional (1D)
potential $V(x)$ by which one
can generate a one-parameter family of isospectral potentials
$V_{iso}(x;\lambda)$, where $\lambda$ is a real, labeling parameter of each
member potential in the set. As a matter of fact, Khare and Sukhatme \cite{ks}
were also able to construct multiparameter families of isospectral
potentials, but we shall be concerned only with the one-parameter case
in the following.
SIDDM is the physical formulation of a mathematical scheme based on the
general Riccati solution, which has been introduced by Mielnik \cite{mi}.
The strictly isospectral techniques are well
established for nodeless bound states, resulting from any pair combination
of the well-known Abraham-Moses procedure, the Pursey one,
and the ``supersymmetric" Darboux one \cite{ks}. Recently, Sukhatme and
collaborators \cite{suk} extended the usage of the double Darboux
construction to energy states in the continuum region
of the spectrum and obtained several families of isospectral potentials with
bound states in the continuum. In particular, they performed the double
Darboux
construction on the half line free particle wavefunction $u_{0}=\sin kr$,
obtaining bound states embedded in the continuum for this simple
but relevant case.

Let us take a simple 1D Schr\"odinger equation for a
wavefunction $u(x)$ at an arbitrary energy level $E$ in the
standard form $-u^{''}(x)+V_{b}(x)u(x)=Eu(x)$.
Then, the strictly isospectral potentials with respect to $V_{b}$
obtained by means of SIDDM, by which one delets and reinserts the level $E$
in the spectrum will be \cite{ks}
$$
V_{iso}(x;\lambda)=V_{b}(x)-2\frac{d^2}{dx^2}[\ln({\cal I}+\lambda)]~,
\eqno(1)
$$
where ${\cal I}(x)=\int _{-\infty}^{x}u^2(y)dy$ and
$\lambda$ is the so called isospectral parameter,
which is a real quantity. The isospectral family of potentials should be
understood in the sense that the whole family has the same supersymmetric
partner ``fermionic" potential given by
$V_{f}(x)=V_{b}(x)-2\frac{d^2}{dx^2}(\ln u)$.
When $u$ is a nodeless state there are no difficulties in applying the
double Darboux technique. However, when one uses an eigenfunction with nodes,
the corresponding ``fermionic" potential will have singularities.
Nevertheless, the resulting isospectral family of
potentials is free of singularities \cite{suk} and this makes the method
viable. Another basic result of the scheme is that the $u$ wavefunction is
changed into $u/[{\cal I}(x)+\lambda]$, i.e., a spatial damping is introduced.


Let us pass now to the quantum Taub model which has been studied in some
detail by Ryan and collaborators \cite{ry} who found that it is separable.
Indeed, the Taub Wheeler-DeWitt equation is
$$
\frac{\partial ^2\Psi}{\partial \alpha ^2}-
\frac{\partial ^2\Psi}{\partial \beta ^2}+e^{4\alpha}V(\beta)\Psi=0~,
\eqno(2)
$$
where $V(\beta)=\frac{1}{3}(e^{-8\beta}-4e^{-2\beta})$. Eq.~(2)
can be separated in the variables
$x_1=4\alpha-8\beta$ and $x_2=4\alpha -2\beta$.
Thus, one gets two independent
1D problems for which the supersymmetric
procedures are standard practice.
The two equations are as follows
$$
-\frac{\partial ^2 u_1}{\partial x_1^2}+\frac{1}{144}e^{x_1}u_1=
\frac{\omega ^2}{4}u_1
\eqno(3)
$$
and
$$
-\frac{\partial ^2 u_2}{\partial x_2^2}+\frac{1}{9}e^{x_2}u_2=
\omega ^2 u_2~,
\eqno(4)
$$
where we have already multiplied both sides in Eqs. (3) and (4) by
$-1$ in order to get standard Schr\"odinger equations.
The quantity $\omega$ is mathematically the separation constant, which
physically is related to the wavenumber of a positive energy level.

Mart\'{\i}nez and Ryan have considered a wavepacket solution made of
wavefunctions $\Psi$ having
the form of a product of modified Bessel functions of imaginary order.
We shall slightly modify their $\Psi$ as follows
$$
\Psi\equiv u_1u_2=K_{i\omega}(\frac{1}{6}e^{x_1/2})
[L_{2i\omega}(\frac{2}{3}e^{x_2/2})+
K_{2i\omega}(\frac{2}{3}e^{x_2/2})]
\eqno(5)
$$
since, according to Dunster \cite{du}, the $L$ function defined as
$$
L_{2i\omega}=\frac{\pi i}{2{\rm sinh}
(2\omega \pi)}(I_{2i\omega}+I_{-2i\omega})~,
\eqno(6)
$$
contrary to the $I_{2i\omega}$ function used by Mart\'{\i}nez and Ryan,
being real on the real axis is a better companion for the $K$ function of
imaginary order which is also real on the real axis. In order to introduce
such a change, Dunster invoked the
criteria on the choice of standard solutions for a homogeneous linear
differential equation of the second order due to Miller \cite{Mi}.
In the following,
we shall make the double Darboux construction on the base of
a $\Psi$ wavefunction of the type given in Eq. (5).

The isospectral potential for the $x_1$ variable will be
$$
V_1(x_1;\lambda _1)=\frac{1}{144}e^{x_1}-2\frac{d^2}{dx_1^2}
\ln[\lambda _1 +{\cal I} (x_1)]
\eqno(7)
$$
and for the $x_2$ one
$$
V_2(x_2;\lambda _2)=\frac{1}{9}e^{x_2}-2\frac{d^2}{dx_2^2}
\ln[\lambda _2 +{\cal I} (x_2)]~,
\eqno(8)
$$
where the integrals are
$$
{\cal I} (x_1)=\int_{-\infty}^{x_1}K_{i\omega}^{2}(\frac{1}{6}e^{y/2})dy
\eqno(9)
$$
and
$$
{\cal I} (x_2)=\int_{-\infty}^{x_2}[L_{2i\omega}(\frac{2}{3}e^{y/2})+
K_{2i\omega}(\frac{2}{3}e^{y/2})]^2 dy~.
\eqno(10)
$$
The total Taub isospectral wavefunction has the following form
$$
\Psi ^{T}_{iso}(x_1, x_2;\lambda _1, \lambda _2)\equiv u_{iso, 1}u_{iso, 2}=
\frac{K_{i\omega}(\frac{1}{6}e^{\frac{1}{2}x_1})}{[{\cal I} (x_1)+\lambda _1]}
\frac{L_{2i\omega}(\frac{2}{3}e^{\frac{1}{2}x_2})+
K_{2i\omega}(\frac{2}{3}e^{\frac{1}{2}x_2})}
{[{\cal I} (x_2)+\lambda _2]}~.
\eqno(11)
$$
We have plotted the Taub isospectral potentials and the corresponding
isospectral wavefunction for several values of the parameters
in Figs. 1, 2, 3, 
respectively.
We confirm the previous findings on the strictly isospectral supersymmetric
effects \cite{suk}, except for the fact that the wavefunctions are still
not normalizable (i.e., they are not bound states in the continuum).

In conclusion, SIDDM allows the introduction
of a one-parameter family of isospectral quantum Taub potentials
having more attenuated states in the continuum region of the
spectrum in comparison to the original potential.

The double Darboux method appears to be a quite general and useful method
to generate new sets of quantum cosmological solutions. This is so because
any potential in the Schr\"odinger equation has a classical continuum of
positive energy nonnormalizable solutions. Ryan and collaborators \cite{ry}
were among the first to pay attention to the continuum part of the
Wheeler-DeWitt spectrum. Selecting by means of some
preliminary physical arguments that are of quantum scattering type
one of the continuum solutions \cite{ry},
one can perform the double Darboux construction on that state
and generate by this means strictly isospectral families of
cosmological potentials as well as isospectral cosmological wavefunctions.
The quantum Taub model just illustrates this nice feature of
the double Darboux method.
If one pushes further the picture,
one might say that the incipient universes were nothing else but sets of
strictly isospectral states of a quantum continuum.
The parameter $\lambda$ looks like a decoherence parameter
embodying a sort of ``quantum" cosmological dissipation (or damping) distance.
Finally, one can easily apply the other procedures of
deleting and reinserting energy levels, i.e., combinations of any pairs of
Abraham-Moses procedure, Pursey's one, and the Darboux one \cite{ks}.
However, only the double Darboux method used here leads to reflection and
transmission amplitudes identical to those of the original potential.

\bigskip
\bigskip
This work was partially supported by the Consejo Nacional de Ciencia y
Tecnolog\'{\i}a (CONACyT) (Mexico) Projects 4868-E9406 and 3898P-E9608.
We thank M.P. Ryan for drawing our attention to the quantum Taub model.

\bigskip
\bigskip

{\bf Appendix}

We quote here some important properties
of the modified Bessel functions of imaginary order \cite{du}.

Their behavior at $x\rightarrow 0^{+}$ is as follows
$$
K_{i\omega}(x)=-\Bigg[\frac{\pi}{\omega \sinh(\omega \pi)}\Bigg] ^{1/2}
[\sin(\omega \ln(x/2)-\phi)+O(x^2)]
\eqno(A1)
$$
and
$$
L_{i\omega}(x)=-\Bigg[\frac{\pi}{\omega \sinh(\omega \pi)}\Bigg] ^{1/2}
[\cos(\omega \ln(x/2)-\phi)+O(x^2)]
\eqno(A2)
$$
where the phase $\phi ={\rm arg}[\Gamma (1+i\omega)]$.
The amplitudes of oscillation of the two functions become unbounded as
$\omega\rightarrow 0$ in the neighborhood of the origin. In the
paper we worked at fixed, small positive $\omega$ for computational
reasons.

In the complex plane, as $z\rightarrow \infty$, the asymptotics are
the following
$$
K_{i\omega}(z)=\Bigg[\frac{\pi}{2z}\Bigg] ^{1/2}
e^{-z}[1+0(1/z)], \;\;\;|{\rm arg z}|
\leq 3\pi/2-\delta
\eqno(A3)
$$
and
$$
L_{i\omega}(z)=\frac{1}{\sinh (\omega \pi)}
\Bigg[\frac{\pi}{2z}\Bigg] ^{1/2}e^{z}[1+0(1/z)], \;\;\;|{\rm arg z}|
\leq \pi/2-\delta
\eqno(A4)
$$
where $\delta$ is an arbitrary small positive constant.

As for the zeros of these functions, it is known that $K_{i\omega}(x)$ has
an infinite number of simple positive zeros in $0<x<\omega$ and no zeros
in $\omega\leq x<\infty$, whereas $L_{i\omega}(x)$ has an infinite number of
simple positive zeros. Up to $x=\omega$ the two sets of zeros are interlaced.


\newpage
\centerline{
\epsfxsize=280pt
\epsfbox{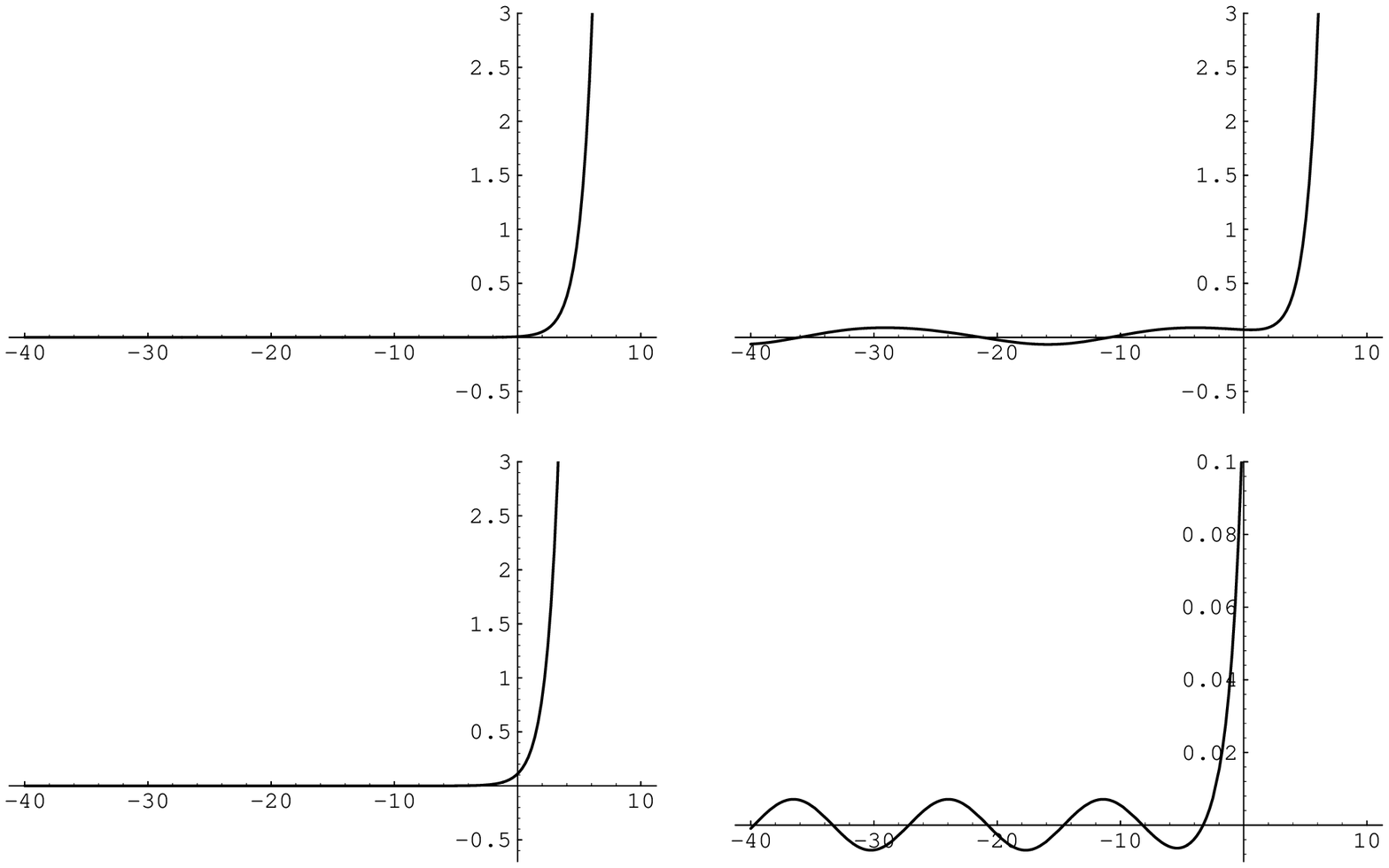}}
\vskip 4ex
\begin{center}
{\small {Fig. 1}\\
The original potentials $V_1(x_1)$ and $V_2(x_2)$ (left side) and the 
corresponding isospectral members for $\lambda_1=\lambda_2=1$ (right 
side) and $\omega=1/4$.}
\end{center}

\centerline{
\epsfxsize=280pt
\epsfbox{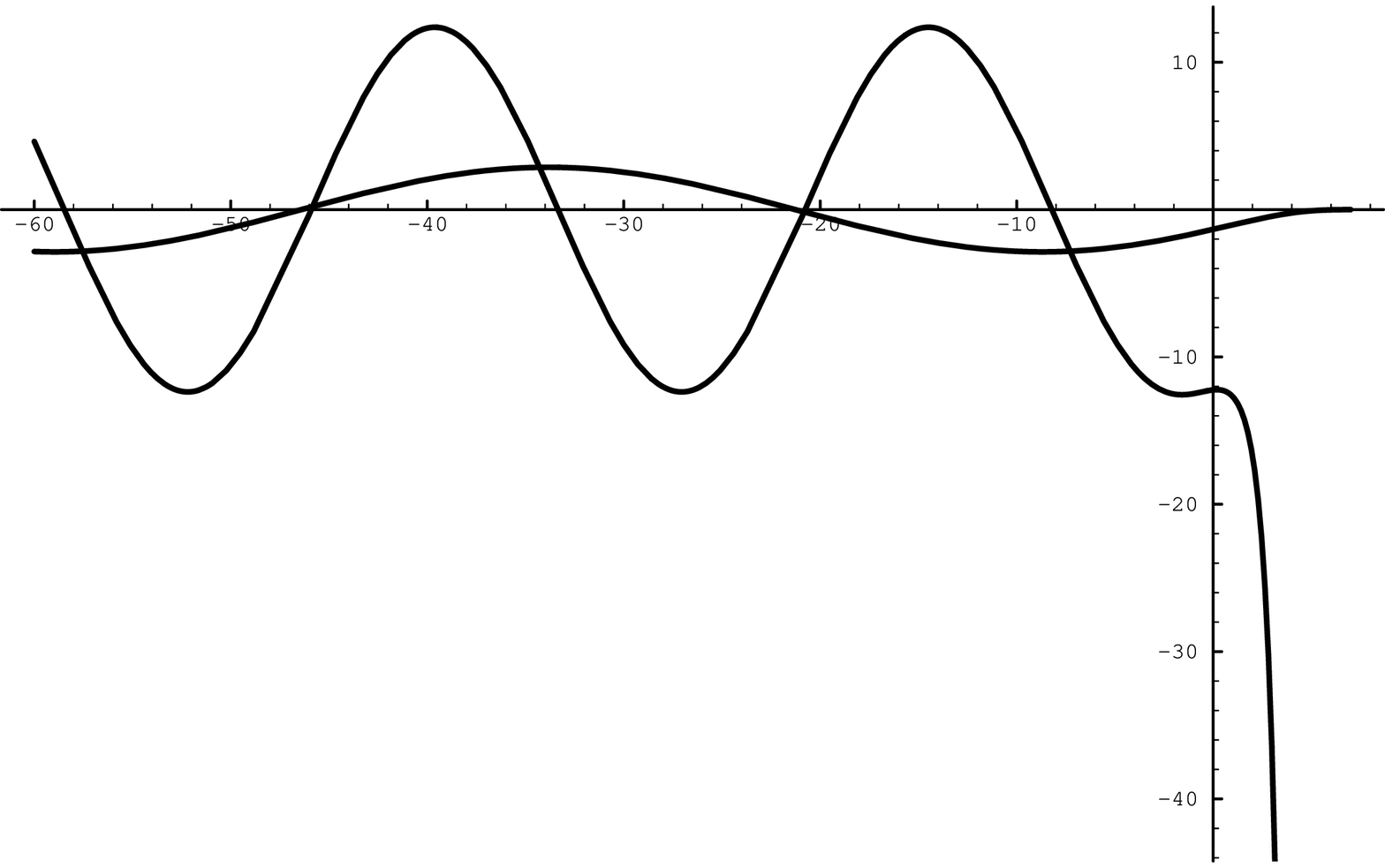}}
\vskip 4ex
\begin{center}
{\small {Fig. 2}\\
The wave functions $u_{ 1}$ (the smaller amplitude one)
and $u_{ 2}$ (the larger amplitude one) of the original
Taub universe for $\omega=1/4$.}
\end{center}

\centerline{
\epsfxsize=280pt
\epsfbox{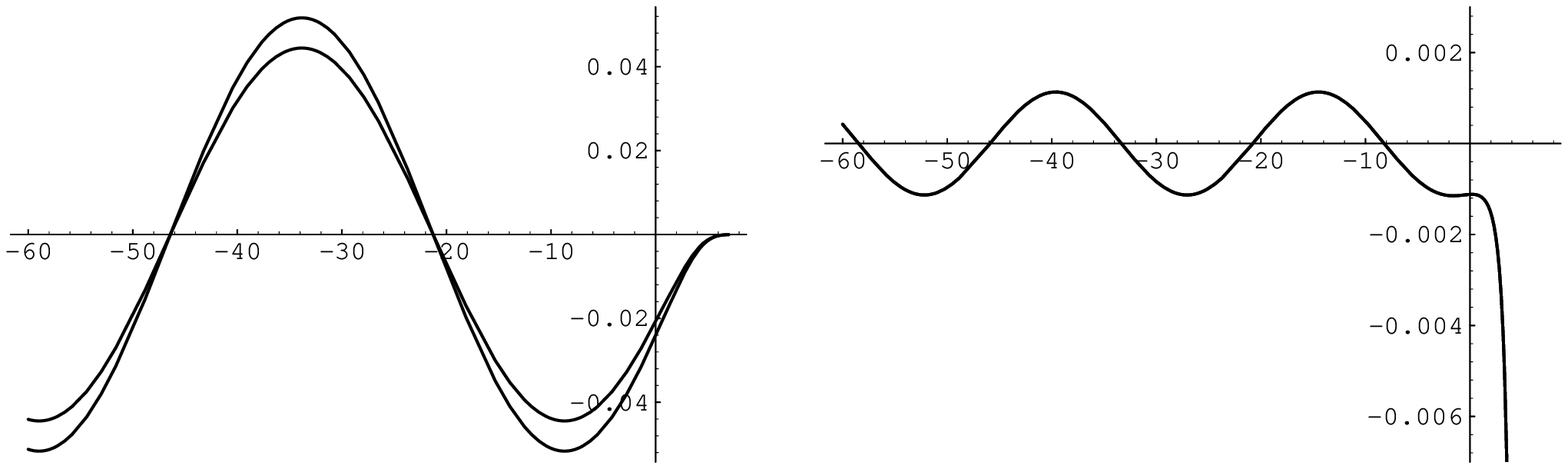}}
\vskip 4ex
\begin{center}
{\small {Fig. 3}\\
 Isospectral wavefunctions $u_{iso,1}$ for  $\lambda_1= 1$ (larger 
amplitude) and $\lambda_1=10$ (smaller amplitude); $u_{iso,2}$ 
for  $\lambda _{2}=1$, respectively, and $\omega= 1/4$.}
\end{center}

\centerline{
\epsfxsize=280pt
\epsfbox{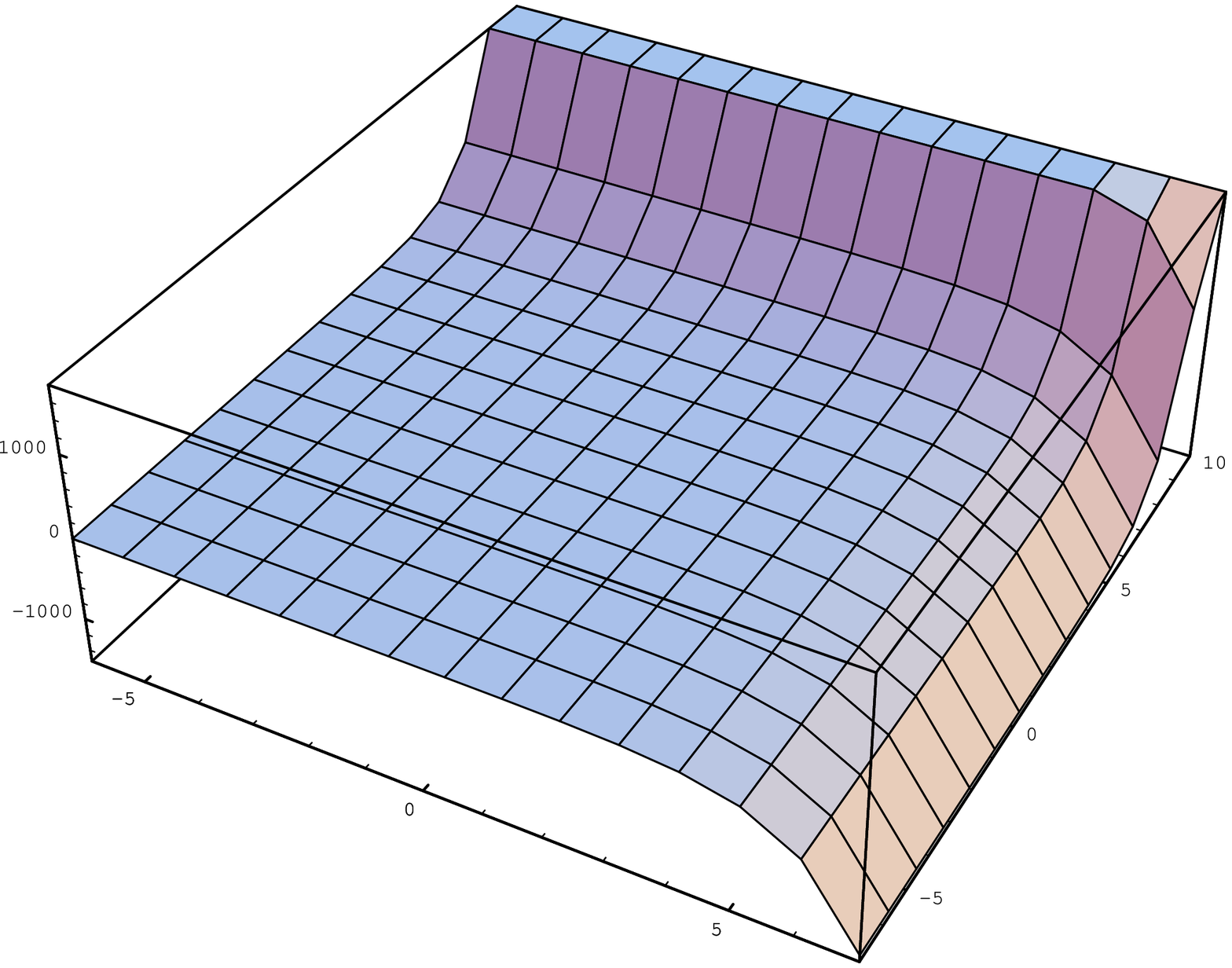}}
\vskip 4ex
\begin{center}
{\small {Fig. 4}\\
 The original Taub potential in the $x_1$, $x_2$ variables.}
\end{center}

\end{document}